\documentclass[prd,twocolumn,preprintnumbers,amsmath,amssymb,nofootinbib,floatfix]{revtex4}

\usepackage{graphicx,bm}

\makeatletter
\def\graphicscale{\twocolumn@sw{0.3}{0.4}}
\def\graphicthreescale{\twocolumn@sw{0.3}{0.4}}

\begin{document}

\title{On the effective Lagrangian of $CP^{N-1}$ models in the large $N$ limit }

\author{Paolo Rossi} 

\address{Dipartimento di Fisica dell'Universit\`a di Pisa,
         Largo Pontecorvo 3, I-56127 Pisa, Italy}

\date{\today}

\begin{abstract}

The effective low energy Lagrangian of $CP^{N-1}$ models in $d < 4$ dimensions can be constructed in the large $N$ limit by solving the saddle point equations in the presence of a constant field strength. The two dimensional case is explicitly worked out and possible applications are briefly discussed.

\end{abstract}


\maketitle



Many properties of quantum field theories can be understood by the evaluation and the study of effective actions depending on the degrees of freedom relevant to the space and energy scales under consideration.
In the low energy limit the approach leading to the construction of the effective potential~\cite{Coleman, Jackiw} has proven to be quite useful in many instances. 

However the extension of the method to gauge theories is not straight-forward, because one cannot choose nontrivial vacuum expectation values for the gauge fields without breaking gauge invariance. A way out involves assuming constant electric and magnetic fields, in which case a gauge invariant effective theory can be constructed. One may then obtain the Euler-Heisenberg Lagrangian~\cite{Heisenberg}, describing the low energy sector of QED.

Here we present a slightly different approach, exploiting the properties of the large $N$ limit. We explore in detail the gauge sector of $CP^{{N-1}}$ models~\cite{Dadda, Witten, Campos1, Campos2, Campos3},  but in principle the method might be extended to large $N$ scalar electrodynamics. 
$CP^{{N-1}}$ models may be interesting for possible applications to condensed matter physics~\cite{Read},  but also because their two-dimensional version shares with QCD many important properties (asymptotic freedom, confinement, $\theta$ dependence) and therefore it may be employed as a laboratory in order to test theoretical ideas that cannot be easily verified in  QCD~\cite{Deldebbio}.

The essential ingredient of our method amounts to recognizing that the saddle point equations describing the model in the large $N$ limit can be explicitly solved also in the presence of space dependent gauge fields, as long as they correspond to constant (gauge-invariant) field strengths.
 Aside from the technicalities of the derivation, the final result is an effective low energy Lagrangian depending only on gauge invariant fields and allowing also for the study of the response to the presence of (gauge invariant) sources.

$CP^{N-1}$ models are defined by the action
\begin{equation}
S_0 (z,{\bar z}) = {N \over 2\,f} \int d^d x\,D_\mu {\bar z} D_\mu z,
\label{eq1}
\end{equation}
where  $z$ is an $N$-component complex vector satisfying ${\bar z}\,z = 1$,  $D_\mu \equiv \partial_\mu + i\,A_\mu$ and $A_\mu \equiv i\, {\bar z} \,\partial_\mu z$. 

One may implement the constraints by introducing Lagrange multiplier fields $\alpha$ and $\lambda_\mu$ and perform the gaussian integration over the unconstrained $z$ fields, obtaining the  effective action
\begin{equation}
S_1 (\lambda_\mu, \alpha) = N\, {\rm Tr \, Ln} [-D_\mu D_\mu +i\,\alpha] -{N \over 2\,f} \int d^d x \,[i\,\alpha],
\label{eq2}
\end{equation}
where now  $D_\mu \equiv \partial_\mu +i\,\lambda_\mu$.

In order to build up generating functionals for the correlations one may add source terms to the action. If one is interested only in the gauge invariant correlations of the gauge fields, the coupling to the source $J_{\mu \nu}$ may take the form
\begin{equation}
N \int d^d x\, F_{\mu \nu} J_{\mu \nu},
\label{eq3}
\end{equation}
where $F_{\mu \nu} \equiv \partial_\mu \lambda_\nu - \partial_\nu \lambda_\mu$, and we introduced a factor $N$ in the definition of the source for the purpose of investigating the large $N$ behavior of the system.

\smallskip
In the large $N$ limit the functional integration over the fields $\alpha$ and $\lambda_\mu$ with a measure determined by the action $S_1$ may be replaced by a saddle point evaluation implying minimization of the action with respect to the fields.

Let's now define the dressed propagator $\Omega (x,y;\alpha,\lambda_\mu)$, solving the equation
\begin{equation}
[-D_\mu D_\mu +i\,\alpha]\,\Omega (x,y;\alpha,\lambda_\mu) = (2\,\pi)^d \delta^{(d)}(x-y),
\label{eq4}
\end{equation}
where we adopted a dimensional regularization in order to take care of singularities that may be present for physical values of $d$.

The saddle point equations for the action $S_1$ in the presence of a (gauge invariant) source are the following:
\begin{eqnarray}
-{i \over N} {\delta S_1 \over \delta \alpha(x)} &\equiv& \Omega (x,x;\alpha,\lambda_\mu)  - {1 \over 2\,f} = 0,\\
{i \over N} {\delta S_1 \over \delta \lambda_\mu (x)} &\equiv& 2\,D_\mu \Omega (x,x;\alpha, \lambda_\mu) = 2\,\partial_\nu J_{\mu \nu}.
\label{eq6}
\end{eqnarray}

The first equation does not depend on $J_{\mu \nu}$ and therefore it can be solved  (in principle) for arbitrary values of $\lambda_\mu(x)$, finding a function $\alpha (\lambda_\mu(x))$. In the logic of the saddle point evaluation of the functional integral, substituting $\alpha (\lambda_\mu)$ in the action is equivalent to performing the integration over the $\alpha$ field. 

The resulting effective action 
$S_\lambda (\lambda_\mu) \equiv  S_1 (\lambda_\mu, \alpha (\lambda_\mu))$
depends only on the gauge field. Therefore the $2n$-point effective vertices appearing in $S_\lambda$ must enjoy explicit gauge invariance.
It is trivial to show that the definition of $\alpha (\lambda_\mu)$ implies also that the value of $\lambda_\mu$ found by minimizing $S_1$ will minimize also $S_\lambda$, because the total variation of $S_\lambda$ with respect to $\lambda_\mu$ coincides with the partial variation of $S_1$ with respect to $\lambda_\mu$.

The function $\alpha(\lambda_\mu)$ can be determined in the form of a series in the powers of $\lambda_\mu$ and  of its derivatives, and it is easy to check that the series is even in $\lambda_\mu$, consistent with the fact that the resulting effective action must not include terms that are odd in the gauge field. 

The coefficients of the series can be interpreted in terms of  Feynman diagrams, and the effect of substituting $\alpha(\lambda_\mu)$ in $S_1$ amounts to combining the bare $2n$ gauge field vertices of $S_1$ with all the tree-level counterterms containing lower order mixed vertices connected only by scalar propagators. This is exactly the procedure needed in order to restore gauge invariance in the ``dressed'' (or ``subtracted'') gauge field vertices that will appear in $S_\lambda$.
As a byproduct of this approach, the combinatorics of the "dressing" may be extracted from the perturbative expansion of the first saddle point equation. 

In turn solving the saddle point equation for the action $S_\lambda$ in the presence of a source  and substituting the result in $S_\lambda$ corresponds to building up the generating function for the gauge invariant (connected) correlations of the gauge fields by summing up all the contributions coming from the tree approximation of the effective theory.

Because of gauge invariance the effective action cannot have a nontrivial dependence from a constant field $\lambda_\mu$, and it must be a function of the field strength and of its derivatives. 
The dressed vertices appearing in the effective action must be transverse and therefore their low momentum behavior must be described by  homogeneous $2n$-degree polynomials in the momenta.  In particular their two dimensional tensor structure is bound to assume the form 
\begin{equation}
\prod_{i=1}^{2n} \,\varepsilon_{\mu_i \nu_i} p_{\nu_i}^{(i)},
\label{eq7}
\end{equation}
with appropriate generalizations to higher dimensions, holding for all $d<4$.

Similar arguments may be applied to the function $\alpha(\lambda_\mu)$ finding that it must be gauge invariant and in the low energy limit it must be a (space independent) function of (constant) $F_{\mu \nu}$.

Therefore in the zero-field strength limit $\lambda_\mu$ may be set equal to zero and the first saddle point equation becomes
\begin{eqnarray}
&&\Omega(x,x;\alpha_0, 0) = \int {d^d p \over (2\pi)^d} {1 \over p^2 + i\,  \alpha_0}  \nonumber\\
\label{eq8}
 &&={\Gamma(1-{d \over 2})\over (4 \pi)^{d \over 2}} [i\,  \alpha_0]^{ {d \over 2}-1 } = {1 \over 2\,f},
\end{eqnarray}
allowing for the elimination of the coupling $f$ in favor of the mass scale $m^2 \equiv i\,\alpha_0$.

The direct calculation of the low energy limit of $\alpha(\lambda_\mu)$ for small field strength allows to verify that
\begin{equation}
i\, \alpha(F_{\mu \nu}) = m^2 + \bigl({d \over 2}-2 \bigr) {F_{\mu \nu} F_{\mu \nu} \over 12\,m^2 } + O (F_{\mu \nu}^4).
\label{eq9}
\end{equation}
In order to construct the gauge invariant effective potential to all orders in the field strength we are left with the task of extracting the function $\alpha (F_{\mu \nu})$ from the solution of the saddle point equation
\begin{equation}
\Omega (x,x;\alpha,\lambda_\mu) = \Omega(x,x, \alpha_0,0).
\label{eq10}
\end{equation}

However  in the case of  constant $F_{\mu \nu}$ we may assume $\lambda_\mu = {1 \over 2} F_{\mu \nu} x_\nu$ (in the transverse gauge) and a significant simplification occurs in the evaluation of $\Omega$ since
\begin{equation}
\Omega^{-1} = -\partial_\mu \partial_\mu + F_{\mu \nu} \,L_{\mu \nu} +{1 \over 4} F_{\mu \rho} F_{\nu \rho} \,x_\mu x_\nu + i\,\alpha,
\label{eq11}
\end{equation}
where $L_{\mu \nu} = i\,(x_\mu \partial_\nu - x_\nu \partial_\mu)$ is the angular momentum operator.

In order to exploit dimensional regularization we may decompose the space degrees of freedom noticing that for $d<4$ the tensor $ F_{\mu \rho} F_{\nu \rho}$ is a projection operator on a two-dimensional subspace. Hence it is possible to write
\begin{equation}
\Omega^{-1} = \bigl[p_1^2 + p_2^2 + B\,(x_1\,p_2 - x_2\,p_1) +{1 \over 4} B^2 (x_1^2 +x_2^2) \bigr] + A + p_\perp^2,
\label{eq12}
\end{equation}
where $B$ is the modulus of the nontrivial eigenvalues of $F_{\mu \nu}$, $A = i\,\alpha$ and $p^\perp_\mu$ are the components of momentum orthogonal to the $(x_1,x_2)$ plane.

We may now define the (creation and annihilation) operators
\begin{eqnarray}
a_\pm \equiv p_1 \pm i\,p_2 + {i \over 2} B (x_1 \pm i\,x_2),\nonumber\\
a_\pm^\dagger \equiv p_1 \mp i\,p_2 - {i \over 2} B (x_1 \mp i\,x_2),
\label{eq13}
\end{eqnarray}
and notice that it is possible to write
\begin{equation}
\Omega^{-1} = a_\pm^\dagger a_\pm + B + A + p_\perp^2.
\label{eq14}
\end{equation}

As a consequence $\Omega^{-1}$ can be interpreted as an Hamiltonian whose eigenstates are the eigenstates of energy and angular momentum of a two dimensional quantum harmonic oscillator, and $\Omega$ is the corresponding Green's function.
Hence we can formally write
\begin{eqnarray}
 \Omega(x,y) &=&\int {d^{d-2} p_\perp \over (2 \pi)^{d-2}} \, e^{i p_\mu^\perp (x_\mu^\perp-y_\mu^\perp)}\nonumber\\
&& \times \sum_{n,m} {\psi_{nm}(x_1,x_2)| \psi^*_{nm}(y_1,y_2)\over (n\pm m+1)B + A+ p_\perp^2}.
\label{eq15}
\end{eqnarray}
and  in particular
\begin{eqnarray}
&& \Omega(x,x) = \int {d^{d-2} p_\perp \over (2 \pi)^{d-2}} \sum_{n,m} {|\psi_{nm}(x_1,x_2)|^2 \over (n\pm m+1)B + A+ p_\perp^2}\nonumber\\
&&={\Gamma(2-{d \over 2})\over (4 \pi)^{{d \over 2}-1}} \sum_{n,m} \bigl[(n \pm m+1) B +A \bigr]^{{d \over 2} -2} |\psi_{nm}|^2.
\label{eq16}
\end{eqnarray}

Moreover, since the only excitations of the ``Hamiltonian'' are left (or right) circular quanta, we can take advantage of this fact and express the result in terms of the wavefunctions $\psi_{n_l n_r}(x_+,x_-)$, where $x_\pm \equiv x_1\pm i x_2$ and $n_{l,r} = n \pm m$. 

Let's now describe the properties of the wavefunctions $\psi_{n_l n_r}(x_+,x_-)$:
\begin{equation}
\psi_{n_l n_r} (x_+,x_-) = {\beta \over \sqrt{\pi\,n_l!\,n_r!}} P_{n_l n_r} (\beta x_+,\beta x_-) e^{-{1 \over 2} \beta^2 x_+ x_-},
\label{eq17}
\end{equation}
where $\beta \equiv \sqrt{B/2}$ and we introduced the polynomials
\begin{equation}
P_{ab}(s_+,s_-) \equiv \sum_l {(-1)^l \over l!} {a! \,(s_+)^{a-l} \over (a-l)!}{b!\,(s_-)^{b-l}\over (b-l)!}.
\label{eq18}
\end{equation}

It is possible to prove the following general identity:
\begin{eqnarray}
&&\sum_n {1 \over n!} P_{an}(s_+,s_-) \,P^*_{bn} (t_+,t_-) = \nonumber\\
&& =(-1)^b P_{ab}(t_+-s_+,t_--s_-) e^{t_+s_-},
\label{eq19}
\end{eqnarray}
and as a special case, performing the summation over the right circular quanta
\begin{eqnarray}
&& \sum_{n_r} \psi_{n_l n_r}(x_+,x_-) \,\psi_{n_l n_r}(y_+,y_-) = e^{{1 \over 2}\beta^2 (x_-y_+ - x_+y_-)}\nonumber\\
&& \times (-1)^{n_l} {\beta \over \sqrt{\pi}} \psi_{n_l n_l}(y_+-x_+,y_--x_-) ,
\label{eq20}
\end{eqnarray}
where the exponent in the r.h.s is purely imaginary.

As a consequence we  obtain
\begin{eqnarray}
\Omega(x,y) &=&\int {d^{d-2} p_\perp \over (2 \pi)^{d-2}} \, e^{i p_\mu^\perp (x_\mu^\perp-y_\mu^\perp)} \sqrt{B \over 2 \pi} e^{i{B \over 2}(x_1 y_2 -x_2 y_1)} \nonumber \\
&&\times  \sum_{n_l} (-1)^{n_l}  {\psi_{n_l n_l}(y_+-x_+, y_--x_-) \over (2\,n_l+1)B + A+ p_\perp^2},
\label{eq21}
\end{eqnarray}
 where $\psi_{n_l n_l}$ actually depends only on the combination $ (y_+-x_+)(y_--x_-) \equiv [(y_1-x_1)^2 + (y_2-x_2)^2]$.

It is worth noticing that the space dependence of $\Omega(x,y)$ is dictated by translation invariance, rotation invariance and gauge covariance. 

In particular we may exploit the translation invariance of the problem and recognize that in the case of constant field strength a translation amounts to a gauge transformation, and it is possible to find a translation such that $\Omega^{-1}$ becomes explicitly translation invariant and the phase in $\Omega$ is absorbed, making the new $\Omega$ explicitly translation invariant.
Moreover because of rotation invariance the term depending on the angular momentum $L_{\mu \nu}$ in the new $\Omega^{-1}$ can be set equal to $0$.

In any case, observing that $\psi_{n_l n_l} (0,0)= (-1)^{n_l} \sqrt{B \over \ 2\pi}$, we immediately obtain the relationship
\begin{equation}
\Omega(x,x) = {\Gamma(2-{d \over 2}) \over (4\pi)^{d \over 2}} \sum_{n_l} 2\,B [(2\,n_l+1)B + A]^{{d \over 2}-2}.
\label{eq22}
\end{equation}

This expression is formally divergent for physical values of $d$ and it will need further regularization.

To this purpose the expression for $\Omega(x,x,A,B)$ can be rephrased in the form~\cite{Slobo}
\begin{eqnarray}
 \Omega (x,x;A,B) &=&\int_0^\infty {dz \over (4\pi)^{d \over 2} }\,z^{1- {d \over 2}} \sum_{n=1}^\infty 2B\,e^{-z(A+B+2\,n\,B)}  \nonumber\\
&=& \int_0^\infty {dz \over (4\pi)^{d \over 2} }\,z^{- {d \over 2}} {z B \over \sinh z B}  e^{-z\,A}.
\label{eq23}
\end{eqnarray}

Noticing that
\begin{equation}
\Omega (x,x;m^2,0) = \int_0^\infty {dz \over (4\pi)^{d \over 2}}\,z^{- {d \over 2}}   e^{-z\,m^2},
\label{eq24}
\end{equation}
it is now easy to regularize the above expression by writing
\begin{eqnarray}
&&\Omega(x,x;A,B)-\Omega(x,x,m^2,0) = \nonumber\\
&&=\int_0^\infty {dz \over (4\,\pi)^{d \over 2} } \,z^{- {d \over 2}} \Bigl[ {z B \over \sinh z B}  e^{-z\,A}- e^{-z\,m^2}\Bigr].\qquad
\label{eq25}
\end{eqnarray}

The l.h.s of this equation is finite for all $d < 4$, and by  using the expansion 
\begin{equation}
 {x \over \sinh x} = 1+\sum_{n=1}^\infty {(2- 2^{2n}) B_{2n}  \over (2n)!} x^{2n} \equiv \sum_{n=0}^\infty {C_n \over (2n)!} x^{2n},
\label{eq26}
\end{equation}
where $B_{2n}$ are the Bernoulli numbers, changing variables to $zA = t$ and performing some trivial integration one finds the asymptotic series representation
\begin{eqnarray}
&&\Omega(x,x;A,B)-\Omega(x,x,m^2,0) = \nonumber \\
&& ={1 \over (4\,\pi)^{d \over 2} } A^{{d \over 2}-1} \Bigl[ \Gamma (1-{d \over 2}) \Bigl(1 - \bigl({A \over m^2}\bigr)^{1-{d \over 2}} \Bigr)\nonumber\\
&&\qquad \qquad+ \sum_{n=1}^\infty{\Gamma(2n+1-{d \over 2}) \over \Gamma(2n+1)} C_n \bigl({B \over A}\bigr)^{2n}  \Bigr].\qquad
\label{eq27}
\end{eqnarray}

In order to solve the first saddle point  equation and find $A(B^2)$ the r.h.s. of the above expression must be set equal to zero.

In the lowest order of the expansion one may reproduce the above mentioned result
\begin{equation}
A(B^2) \approx m^2 - {1 \over 6} \bigl(2-{d \over 2} \bigr){ B^2 \over m^2} + O (B^4).
\label{eq28}
\end{equation}

It is also possible to confirm the perturbative result, holding in the limit when $d \rightarrow 4$ from below,
\begin{equation}
A(B^2) \rightarrow m^2 + O (4-d).
\label{eq29}
\end{equation}

A representation of the effective potential in $d$ dimensions may be obtained from the integration of the above result:
\begin{equation}
V(A,B)-V(A,0) = - \bigl({A \over 4\,\pi}\bigr)^{d \over 2} \sum_{n=1}^\infty {\Gamma(2n-{d \over 2}) \over \Gamma(2n+1)} C_n \bigl({B \over A}\bigr)^{2n}.
\label{eq30}
\end{equation}

\bigskip
In two dimensions one may follow a different strategy in the evaluation of the integrals, by writing
\begin{eqnarray}
&&\Omega(x,x;A,B)-\Omega(x,x;m^2,0) =\nonumber\\
&&= {1 \over 4\,\pi } \int_0^\infty dz\, \Bigl[ {2\, B\, e^{-z(A+B)} \over 1-e^{-2zB}}  - {e^{-z\,m^2} \over z}\Bigr] \nonumber\\
&&= {1 \over 4\,\pi } \int_0^\infty dt\, \Bigl[ { e^{-t{A+B \over 2B}} \over 1-e^{-t}} -{e^{-t} \over t}\Bigr] + \Bigl[{e^{-t} \over t}- {e^{-t {m^2 \over 2B}} \over t}\Bigr]  \nonumber\\
&&= -{1 \over 4\,\pi} \Bigl[\psi\bigl({1 \over 2}+{A \over 2\,B}\bigr) + \ln {2\,B \over m^2}\Bigr],
\label{eq31}
\end{eqnarray}
where $\psi(z) \equiv {d \over dz} \ln \Gamma(z)$ is the digamma function.

It is now possible to reconstruct the complete effective potential, by integrating the above expression with respect to $A$ and imposing proper boundary conditions in order to determine the residual dependence on $B$. 

The $2$-dimensional result is then
\begin{eqnarray}
&&V = {1 \over 4\,\pi} \Bigl[ B \ln 2\pi- A\,\ln {2 B \over m^2} -2 B\,\ln \Gamma \bigl({1 \over 2} +{A \over 2 B} \bigr) \Bigr] \nonumber\\
&&={A \over 4\,\pi}\Bigl[ 1 - \ln \bigl({A \over m^2}\bigr) + \sum_{n=1}^\infty {C_n \over 2\,n  (1- 2\,n)} \bigl( {B \over A} \bigr)^{2n}\Bigr].
\label{eq32}
\end{eqnarray}

As a consequence we also obtain
\begin{eqnarray}
{\partial V \over \partial A} &=&-{1 \over 4\,\pi} \Bigl[\psi\bigl({1 \over 2}+{A \over 2 B}\bigr) + \ln {2 B \over m^2}\Bigr] \nonumber\\
 &=&\, {1 \over 4\,\pi} \Bigl[\sum_{n=1}^\infty {C_n \over 2\,n} \bigl( {B \over A}\bigr)^{2n} -\ln {A \over m^2}\Bigr],\qquad
\label{eq33}
\end{eqnarray}
\begin{eqnarray}
 {\partial V \over \partial B} &=& {1 \over 4\,\pi} \Bigl[\ln 2 \pi - {A \over B} -2\, \ln  \Gamma \bigl({1 \over 2}
 +{A \over 2 B} \bigr) \nonumber\\
&&+{A \over B} \psi \bigl({1 \over 2} +{A \over 2 B} \bigr)\Bigr] \nonumber\\
&=& {1 \over 4\,\pi} \Bigl[\sum_{n=1}^\infty {C_n \over 1- 2\,n} \bigl( {B \over A} \bigr)^{2n-1}\Bigr].
\label{eq34}
\end{eqnarray}

The saddle point equation ${\partial V \over \partial A}  = 0$ may now be employed in order to find the function $A(B^2)$.
The first few terms in the expansion of ${A \over m^2}$ in powers of ${B \over m^2}$ are
\begin{equation}
{A \over m^2} \approx 1 -{1 \over 6} \bigl({B \over m^2}\bigr)^2+ {3 \over 40} \bigl({B \over m^2}\bigr)^4 + O\Bigl(\bigl({B \over m^2}\bigr)^6\Bigr).
\label{eq35}
\end{equation}

The function $V_\lambda (B) \equiv V(A(B^2),B)$  is an even function of $B$ and it is the generating functional for the low energy vertices of the gauge fields in the large $N$ limit of $CP^{N-1}$ models, thus acting as the two dimensional counterpart of the Euler-Heisenberg Lagrangian. 

The first few terms in the expansion of $V_\lambda (B)$ are
\begin{equation}
V_\lambda (B) \approx {1 \over 4\pi} \bigl[ m^2 + {1 \over 6} {B^2 \over m^2} -{1 \over 40} {B^4 \over m^6} + O\bigl({B^6 \over m^{10}}\bigr) \Bigr]
\label{eq36}
\end{equation}

Another possible application of the above results may consist in the evaluation of the leading order in the $1/N$ expansion of the $\theta$ dependence of the vacuum energy. To this purpose it is necessary to solve the coupled equations
\begin{equation}
{\partial V \over \partial A} = 0, \qquad \qquad {\partial V \over \partial B} = {i \over 2 \pi} {\theta \over N},
\label{eq37}
\end{equation}

In turn this requires a continuation of the solutions  to complex values of the variables $A$ and $B$. 

It is intriguing to notice that in the complexified version of the equations a nonanalytic dependence on ${B \over A}$ becomes apparent. Indeed by assuming $A$ real and $B \equiv i \bar B$ purely imaginary the effective potential takes the form
\begin{eqnarray}
V(A, \bar B) &=& {1 \over 4 \pi} \Bigl( 2\, \bar B \,\Im \bigl[\ln \Gamma ({1 \over 2} + i {A \over 2\bar B})\bigr] -  A \ln {2\bar B \over m^2}\Bigr) \nonumber\\
&+& {i \over 4 \pi} \bar B \, \ln \bigl(1 + e^{-{\pi A \over 2 \beta}} \bigr)
\label{eq38}
\end{eqnarray}
However the saddle point equations may be solved order by order in $ \hat \theta \equiv {\theta \over N}$, assuming ${A \over m^2}$ to be an even function and ${B \over m^2}$ an odd function of $\hat \theta$, in which case the continuation from imaginary to real values of $\hat \theta$ can be performed without any difficulty. The results of this approach will be presented in a forthcoming publication~\cite{Bonati}.

\acknowledgments
I am deeply indebted to M. Mintchev and E. Vicari for many useful discussions and for critical reading of the manuscript.

\end{document}